\documentclass[twocolumn,prl,showpacs,floatfix,amsfonts,superscriptaddress]{revtex4}
  
\usepackage[]{graphicx}  
\usepackage{amsbsy}  
  
\begin{document}  
\title{Inequivalent routes across the Mott transition in V$_{2}$O$_{3}$ explored by X-ray absorption}  
\author{F.~Rodolakis}  
\affiliation{Laboratoire de Physique des Solides, CNRS-UMR 8502, Universit\'{e} Paris-Sud, F-91405 Orsay, France}  
\affiliation{Synchrotron SOLEIL, L'Orme des Merisiers, Saint-Aubin, BP~48, 91192 Gif-sur-Yvette Cedex, France}  
\author{P.~Hansmann}  
\affiliation{Institut for Solid State Physics, Vienna University of Technology, 1040 Vienna, Austria}  
\affiliation{Max-Planck-Institut f\"ur Festk\"orperforschung, Heisenbergstrasse 1, D-70569 Stuttgart, Germany} 
\author{J.-P.~Rueff}  
\affiliation{Synchrotron SOLEIL, L'Orme des Merisiers, Saint-Aubin, BP~48, 91192 Gif-sur-Yvette Cedex, France}  
\affiliation{Laboratoire de Chimie Physique--Mati\`ere et Rayonnement, CNRS-UMR~7614, Universit\'e Pierre et Marie Curie, F-75005 Paris, France}  
\author{A.~Toschi} 
\affiliation{Institut for Solid State Physics, Vienna University of Technology, 1040 Vienna, Austria}  
\author{M.W.~Haverkort}  
\affiliation{Max-Planck-Institut f\"ur Festk\"orperforschung, Heisenbergstrasse 1, D-70569 Stuttgart, Germany}  
\author{G.~Sangiovanni}  
\affiliation{Institut for Solid State Physics, Vienna University of Technology, 1040 Vienna, Austria}  
\author{A.~Tanaka} 
\affiliation{Department of Quantum Matter, ADSM, Hiroshima University, 
  Higashi-Hiroshima 739-8530, Japan} 
\author{T.~Saha-Dasgupta}  
\affiliation{S.N.Bose Center for Basic Sciences, Salt Lake, Kolkata, India} 
\author{O.K.~Andersen}  
\affiliation{Max-Planck-Institut f\"ur Festk\"orperforschung, Heisenbergstrasse 1, D-70569 Stuttgart, Germany} 
\author{K.~Held}  
\affiliation{Institut for Solid State Physics, Vienna University of Technology, 1040 Vienna, Austria}  
\author{M.~Sikora}  
%\altaffiliation[]{Current address: AGH University of Science and Technology, Krakow, %Poland} 
\affiliation{ESRF, 6 rue Jules Horowitz, BP~220, 38043 Grenoble Cedex, France}  
%\author{P.~Glatzel}  
%\affiliation{ESRF, 6 rue Jules Horowitz, BP~220, 38043 Grenoble Cedex, France}  
%\author{J.-L.~Hazemann}  
%\affiliation{ESRF, 6 rue Jules Horowitz, BP~220, 38043 Grenoble Cedex, France}  
\author{I.~Alliot}  
\affiliation{ESRF, 6 rue Jules Horowitz, BP~220, 38043 Grenoble Cedex, France}  
\affiliation{CEA/DSM/INAC/NRS 17 avenue des Martyrs 38000 Grenoble}  
\author{J.-P.~Iti\'{e}}  
\affiliation{Synchrotron SOLEIL, L'Orme des Merisiers, Saint-Aubin, BP~48, 91192 Gif-sur-Yvette Cedex, France}  
\author{F.~Baudelet}  
\affiliation{Synchrotron SOLEIL, L'Orme des Merisiers, Saint-Aubin, BP~48, 91192 Gif-sur-Yvette Cedex, France}  
\author{P.~Wzietek}  
\affiliation{Laboratoire de Physique des Solides, CNRS-UMR 8502, Universit\'{e} Paris-Sud, F-91405 Orsay, France}  
\author{P.~Metcalf}  
\affiliation{Department of Chemistry, Purdue University, West Lafayette, Indiana 47907, USA}  
\author{M.~Marsi}  
\affiliation{Laboratoire de Physique des Solides, CNRS-UMR 8502, Universit\'{e} Paris-Sud, F-91405 Orsay, France}  
  
\date{\today}  
  
\begin{abstract}  
The changes in the electronic structure of V$_2$O$_3$ across the 
metal-insulator transition induced by temperature, doping and pressure 
are identified using high resolution x-ray absorption spectroscopy at 
the V pre $K$-edge. Contrary to what has been taken for granted 
so far, the metallic phase reached under pressure is shown to differ 
from the one obtained by changing doping or temperature. Using a novel 
computational scheme, we relate this effect to the role and occupancy 
of the $a_{1g}$ orbitals. This finding unveils the inequivalence of different routes across the Mott transition in V$_2$O$_3$.
%the existence of a new route through the phase diagram of V$_2$O$_3$.)
\end{abstract}  
  
\maketitle  
  
%%%introduction  
Some materials present metal-insulator transitions (MIT) without any changes in  crystal structure or long-range magnetic order. These phenomena, known as Mott-Hubbard 
transitions, constitute a fundamental signature of strong electronic 
correlations. The physics emerging in the vicinity of these 
transitions is highly non-trivial and the properties of such materials 
depend on small changes in the electronic structure induced by 
external parameters~\cite{Imada1998,Dagotto2005}. Several features of 
the MIT have been successfully clarified by resorting to realistic 
many-body calculations~\cite{Georges1996}. Yet, contrary to common 
assumptions, a growing number of experimental facts are revealing that 
this physical process is also strongly dependent on the route followed 
through the MIT. Here we show this to be the case for vanadium 
sesquioxide (V$_2$O$_3$).  
The isostructural MIT in Cr-doped V$_2$O$_3$ is considered as the 
textbook example of a Mott transition, which occurs between a 
paramagnetic insulator (PI) and a paramagnetic metallic (PM) phase by 
changing doping level ($x$), temperature (T) or pressure 
(P)~\cite{McWhan1973}. As such V$_2$O$_3$ has served as a test bed 
for many theoretical models 
\cite{Castellani1978,Held2001,Elfimov2002a,Keller2004,Poteryaev2007,Laad2006,Tanaka2002} 
and a sustained experimental effort.   
 
Among the different experimental methods recently employed to study 
the electronic properties  of the Mott transition in Cr-doped 
V$_2$O$_3$~\cite{Limelette2003,Mo2003,Mo2006,Rodolakis2009}, X-ray 
absorption spectroscopy (XAS) has played a crucial role. It was the 
detailed investigation of the V $L_{2,3}$ absorption 
edges~\cite{Park2000} that demonstrated the necessity of abandoning 
the simple one band, $S=1/2$, model to obtain a realistic description 
of the changes in the electronic structure at the phase transition. XAS can also be 
performed at the V $K$-edge in the hard x-ray range: in this case, the 
pre-edge will carry most of the physical information we are interested 
in, as it is predominantly due to $1s\rightarrow 3d$ transitions. The 
excitations in this spectral region are influenced by the core hole 
and should be considered to be of an excitonic nature. The advantages
over the $L_{2,3}$ edge studies are (i) a more straightforward  
interpretation (due to the simple symmetry of the $s$-core hole, the 
multiplet structure reveals a direct view on the $d$-states) and (ii) the 
possibility of applying external pressure as diamond anvil cells 
are compatible with hard X-rays.  

We used V $K$-edge XAS to explore extensively the MIT in V$_2$O$_3$ by 
changing temperature, doping and applying an external pressure. The 
pre $K$-edges were analyzed by a novel computational scheme combining 
local density approximation and dynamical mean field theory (LDA+DMFT) \cite{HeldAdvPhys} 
with configuration interaction (CI) full multiplet ligand field 
calculations to interpret subtle differences at the PM-PI 
transition. This allowed us to: (i) observe in detail the changes in 
the electronic excitations, providing also a direct estimate of the Hund's coupling $J$; 
(ii) analyze the physical properties of the PI and PM phase on both 
sides of the MIT, leading to the main result of our work: (iii) 
understand the difference between P, T or doping-induced transitions. 
This difference is mainly related to the occupancy of the $a_{1g}$ orbitals, suggesting 
the existence of a new ``pressure'' pathway between PI and PM in the 
phase diagram.   
 
%%%experimental part  
The experiments were performed on the inelastic x-ray scattering (IXS) 
beamlines ID-26 and BM-30 at the European Synchrotron Radiation 
Facility (ESRF). We used high quality samples of 
(V$_{1-x}$Cr$_x$)$_2$O$_3$ with various doping in the PM ($x=0$) and 
PI phases ($x=0.011$, and 0.028) at ambient conditions. The MIT was  
also crossed for the 0.011 doping by changing temperature and for the 
0.028 doping by pressure. To obtain the best resolution, the XAS 
spectra were acquired in the so-called partial fluorescence yield 
(PFY) mode~\cite{Groot2001}, monitoring the intensity of 
the V-K$\alpha$ 
($2p\rightarrow 1s$) line as the incident energy is swept across the 
absorption edge. Compared to standard XAS, the PFY mode provides 
better resolved absorption spectra as these are partly free from 
core-hole lifetime broadening effect. The gain is particularly  
appreciable at the transition metal $K$ pre-edges as the $1s$ 
core-hole is extremely short lived with respect to the $2p$ core-hole 
and the dipolar tail is strongly reduced. The V-K$\alpha$ line was 
measured with the help of a IXS spectrometer equipped with a Ge(331) 
spherically bent analyzer. For the temperature dependence, the 
crystals were mounted in a cryostat installed on the spectrometer. On 
the other hand, measuring the PFY mode turned out to be difficult under pressure 
because of the weak fluorescence signal in the pressure cell and we 
opted for standard XAS measurements in transmission mode. To maximize 
the throughput, a powder sample ($x=0.028$) was loaded in a diamond 
anvil cell equipped with composite anvils made of a perforated diamond 
capped with a 500-$\mu m$ thin diamond anvil. Pressure was measured 
in-situ by standard ruby fluorescence technique. We used silicon oil 
as a pressure transmitting medium.      
\begin{figure}[htb]  
\includegraphics[width=0.95\linewidth,clip=true]{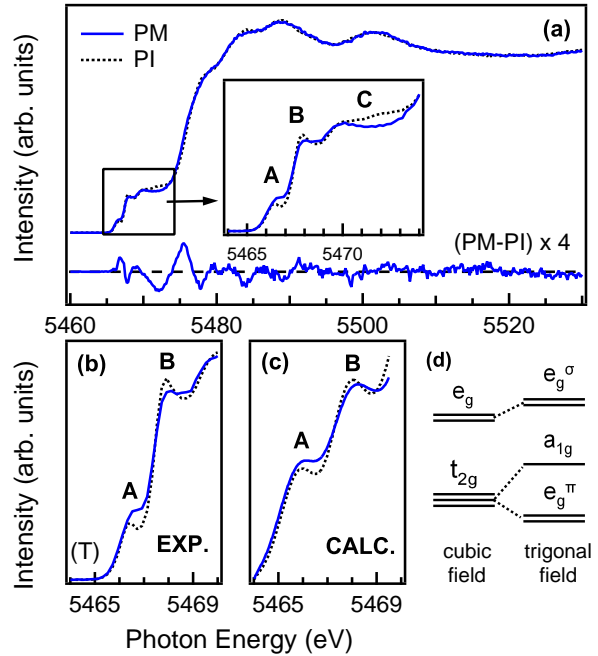}  
\caption{(Color online) (a) V $K$-edge x-ray absorption spectra in 
  (V$_{1-x}$Cr$_x$)$_2$O$_3$, $x=0.011$ powder sample measured as a 
  function of temperature (T) in the PM (200 K, solid line) and PI 
  (300 K, dotted line) phases by PFY XAS; below, PM$-$PI spectral 
  difference; (inset) pre-edge region. (c) Calculated isotropic CI V 
  $K$-edge PM and PI spectra of V$_2$O$_3$. The spectra are 
  compared to the T-dependent experimental data shown in the same 
  energy window (b). (d) From LDA (starting point for our LDA+DMFT and
  CI calculations) we obtain that the cubic part of the ligand field 
  splits the V d-levels in $t_{2g}$ and $e_g$ by $10Dq$ $\sim$2 eV, and that the small 
  trigonal distortion further separates the $t_{2g}$ into $a_{1g}$ and 
  $e^\pi_g$ by $\Delta_{\text{trig}}$ $\sim$0.3 eV.}  
\label{fig:1}  
\end{figure}  
 
The T-dependent absorption spectra are displayed in 
Fig.~\ref{fig:1}(a) for both PM (200 K) and PI (300 K) phases for the 
$x=0.011$, powder sample. The spectra have been normalized to an edge 
jump of unity. We will focus on the pre-edge region, where 
information about the V $d$-states can be easier extracted. It can be 
decomposed into three spectral features (A,B,C) which all vary in 
intensity as the system is driven through the MIT whereas only C is 
considerably broadened (Fig.~\ref{fig:1}(a), inset). Notice that no 
feature is observed below peak A contrary to the early results of 
Ref.~\cite{Bianconi1978} but in agreement with the more recent data of 
Ref.~\cite{Goulon2000}. Within a simplified atomic like picture, one 
could directly relate the intensity of features A,B and C to the 
unoccupied states. Starting from a V-$t_{2g}^2$, $S=1$ configuration, 
one can either add an electron to the $t_{2g}$ subshell (split into one 
a$_{1g}$ and two e$_g^\pi$ states under trigonal distortion of the V sites \cite{paperI}
 as shown in Fig.~\ref{fig:1}(d)) yielding peaks A and B, or add an electron to the $e_g^\sigma$ 
sub-shell which gives rise to the broader peak C. In this picture, 
one could associate peak A with a spin quartet 
%which both correspond to the addition of a $t_{2g}$ electron to the V $3d$ shell 
($S=3/2$) and peak B with a spin doublet ($S=1/2$), split by the Hund's rule exchange.     
%While the polarization dependence of the dipolar part (not shown  
%here) is well explained by LDA calculations~\cite{Meneghini2005}, the  
%pre-edge requires a treatment beyond the standard LDA approach.   

This picture is however oversimplified as the V $d$ electrons are 
{\it strongly correlated} and the spectra are still largely influenced by 
the $1s$ core hole potential. Keeping that in mind, we have simulated 
the pre-edge by combining CI with LDA+DMFT calculations for which the
one particle part (LDA) input corresponds to the level diagram in Fig.~\ref{fig:1}(d). We 
concentrate our analysis to peaks A and B, since peak C relates mainly 
to the unoccupied $e_g^\sigma$ orbitals. These hybridize much stronger 
with the oxygen ligands and thus lack direct information on the 
Mott transition; peak C may also be related to 
non-local excitations (not included here)~\cite{Gougoussis2009} which 
sensitively depend on the metal-ligand distance.
Let us also note that the V sites in V$_2$O$_3$ are non 
centrosymmetric which leads to an on-site mixing of V-$3d$ and
V-$4p$-orbitals and interference between dipole and quadrupole
transitions~\cite{Elfimov2002a} which has been included in our scheme.

%Second, the ground states of the PM 
%and the PI phases starkly differ: as already observed for the undoped 
%compound~\cite{Keller2004,Park2000},
 
Our CI calculations confirm that for the ground state the occupancy ratio between the 
($e_{g}^{\pi}$,$a_{1g}$) and ($e_{g}^{\pi}$,$e_{g}^{\pi}$) states is smaller in 
the PI than in the PM 
phase~\cite{Keller2004,Park2000}: The isotropic 
CI-based calculated XAS spectra in the pre-edge region reported in 
Fig.~\ref{fig:1}(c)
%To help the comparison with the experiment 
%Fig.~\ref{fig:1}(b), we have added the LDA dipolar background (not 
%shown).  
agree well with the experimental data for both  
the energy splitting of features A and B and the ratio of their 
spectral weight (SW) which increases in the PM phase.    
 
\begin{figure}[t!]  
\includegraphics[angle=0,width=0.95\linewidth,clip=true]{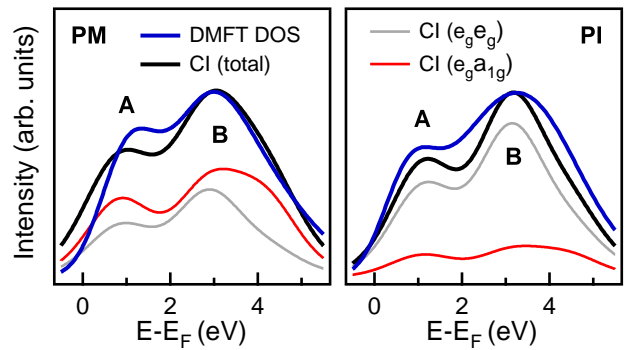}  
\caption{(Color online) Incoherent LDA+DMFT and CI calculations in the 
  pre-edge region; $E_F$ is the Fermi energy. Note the similarity in 
  the main spectral features when crossing the MIT. Also shown are the 
  different  contributions of the CI spectrum labeled accordingly to 
  their initial state: the contribution  of the 
  ($e_{g}^{\pi}$,$a_{1g}$)~$\rightarrow$~($e_{g}^{\pi}$,$e_{g}^{\pi}$,$a_{1g}$) 
  transitions to the peak A(B) is approximately $60\%(55\%)$ in the PM 
  phase and $20\% (15\%)$ in the PI phase.}   
\label{fig:2}   
\end{figure}

Considerable insight can be gained by comparing CI and LDA+DMFT 
calculations. Our LDA+DMFT calculations, performed using 
N$^{th}$-order muffin-tin orbital (NMTO) downfolded Hamiltonian for 
the 1.1\% Cr-doped V$_2$O$_3$ and Hirsch-Fye Quantum Monte Carlo as impurity 
solver, confirm the above mentioned tendency (we obtain a mixing 
of $50$:$50$ and $35$:$65$ for the 
($e_{g}^{\pi}$,$a_{1g}$):($e_{g}^{\pi}$,$e_{g}^{\pi}$) occupation in 
the PM and PI phases). For LDA+DMFT we have  
chosen the same interaction parameter $U=4.2$ eV for the PI phase as 
in Ref.~\cite{Poteryaev2007} and followed their suggestion to slightly 
decrease its value in the PM phase (we assume $U=4.0$ eV)\cite{footnote}. 
Remarkably the simple structure of the core hole 
potential in the $K$-edge spectroscopy allows us to associate the 
pre-edge spectrum with the $k-$integrated spectral function above the 
Fermi energy calculated by LDA+DMFT. The electron--addition part of the
spectral function shows three main features in PM phase: a 
coherent excitation at the Fermi level and a double 
peak associated to the incoherent electronic excitations i.e. the upper Hubbard band (UHB), similarly to 
the undoped compound. In the PI phase obviously, only the latter 
survives. Comparison with the experimental spectra clearly 
shows that the pre-edge features have to be related to the 
``incoherent'' part of the spectral function only. The physical reason is that the 
core hole potential localizes the electrons destroying 
the coherent quasiparticle excitations and making the XAS spectrum 
atomic-like. All the ``incoherent'' LDA+DMFT, CI, and experimental spectra shown in 
Fig.~\ref{fig:2} agree in many aspects, especially as for the splitting of the 
first two peaks by $\approx$2.0 eV ($\approx$1.8 eV in experiment) which 
originates in LDA+DMFT from the Hund's exchange $J$ in the Kanamori 
Hamiltonian. This further validates the choice of $J=0.7$ eV used in our 
calculations contrasting with larger values assumed in previous  
studies~\cite{Keller2004,Laad2006,footnote}, and also clarifies 
the mismatch between XAS and LDA+DMFT spectra reported in the undoped 
V$_2$O$_3$ compound~\cite{Keller2004}. Moreover, the ratio between 
A and B peak displays the same trend in the PM-PI 
transition as the CI (or experimental) data. The quantitative 
difference between the two calculations is attributed to the lack of 
matrix elements in LDA+DMFT.   

The intensity ratio of the first two incoherent excitations peaks A
and B (associated to the quartet and doublet states in the
oversimplified picture) 
thus appears as the key spectral parameter to understand the differences between PM and PI.  
Even in a powder sample, this ratio is still sensitive to the $a_{1g}$ orbital occupation of 
the initial state. Indeed, due to the trigonal distortion a considerable 
spectral weight transfer from the peak B to higher energies 
(corresponding to final states with two $a_{1g}$ electrons in the 
limit of large $\Delta_{trig}$) can take place for the 
($e_g^{\pi},a_{1g}$)  but not for the 
($e_g^{\pi},e_g^{\pi}$) initial state. Therefore, the $K$ pre-edge XAS 
can serve as a direct probe of the $a_{1g}$ orbital occupation in the 
ground-state. As a rule of thumb, the larger the ratio between the SW 
of A and B, the larger the $a_{1g}$ orbital occupation.  
 
\begin{figure}[t!]  
\includegraphics[width=0.95\linewidth]{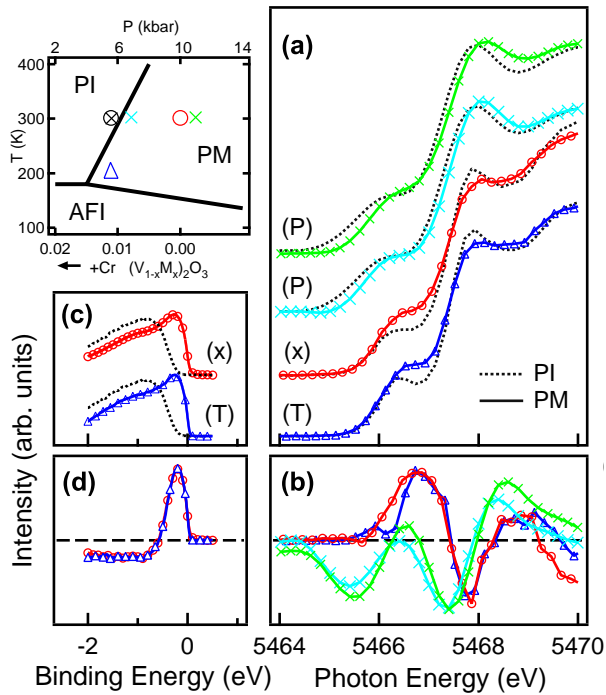}  
\caption{(Color online) (a) V-K XAS spectra for powder samples of 
  (V$_{1-x}$Cr$_x$)$_2$O$_3$ as a function of pressure ($\times$) [$x=0.028$; 
  5 and 7 kbar (lower curves); 5 and 11 kbar (upper curves)] (P),
  temperature ($\triangle$) [$x=0.011$; 200,300 K] (T), and doping 
($\circ$)  [$x=0$, 0.011] ($x$) (cf.\ points in the phase diagram; the pressure 
  scale refers to the $x=0.028$ doping). The spectral differences 
  PM$-$PI ($\times$3) shown for the three datasets (b) demonstrate the 
  nonequivalence between $P$ and temperature-doping. The $x$-$T$ 
  equivalence is confirmed by the photoemission spectra ($h\nu=9$ eV) 
  (c) and their spectral differences (d) (from 
  Ref.~\cite{Rodolakis2009}).}     
\label{fig:3}   
\end{figure}

Hard X-ray absorption also provides a unique way to explore the 
``pressure'' pathway across the MIT, 
%%% 
from which relevant information can be extracted by applying our interpretation scheme.  
%%% 
Fig.~\ref{fig:3}(a) shows the XAS 
powder spectra of the P-induced MIT with the corresponding 
spectra for the T- and doping-driven transition (cf.\ the loci in the 
phase diagram, Fig.~\ref{fig:3}). To ease the comparison, the pressure dataset (P) 
obtained in transmission mode has been deconvolved from the $1s$ Lorentzian 
lifetime broadening (1 eV FWHM) using the GNXAS 
code~\cite{Filipponi2000} to match the improved resolution of 
datasets ($x$) and (T) measured by PFY-XAS.  
%%% 
Together with the PM$-$PI differences in Fig.~\ref{fig:3}(b),
Fig.~\ref{fig:3}(a) clearly evidences that, besides a rigid shift of the 
first two peaks of $\sim$+0.13 eV and contrary to the doping- or T-driven
transition, very small 
changes in spectral shapes and weights are observed in the P-driven MIT.  
%%%% 
In the light of the arguments discussed 
above, our finding proves that the metallic state reached by applying 
pressure is characterized by a much lower occupation of the $a_{1g}$ 
orbitals compared to the metallic state reached just by changing 
temperature or doping. Importantly, the spectra measured 
through the doping induced MIT are identical within the experimental 
uncertainty to those measured through the T-driven transition. The 
temperature-doping equivalence is confirmed by our photoemission data 
displayed in Fig.~\ref{fig:3}(c,d)~\cite{Rodolakis2009} and is consistent with the very similar    
lattice parameter changes across the transition~\cite{McWhan1970}. 
  %observed in the region $200$ 
  %K$<T<400$ K and 
  %$0<x<0.028$~\cite{McWhan1970}. %\footnote{The 
  %angle resolved spectra shown here 
  %were taken at 9 eV %photon energy at normal emission, probing the point in k-space where 
  %the QP peak is stronger. The similarity of the signals from two 
  %different samples is particularly remarkable considering the 
  %sensitivity of the QP photoemission signal to any surface related 
  %perturbation.}.   
The $x$ and $T$ equivalence is also borne out by the observation from 
XAS at the $L$-edges in doped V$_2$O$_3$~\cite{Park2000} that the 
$a_{1g}$ occupation within both the PM or PI phases is mostly 
independent of the doping level. Hence, the local 
incoherent excitations probed by XAS at the V $L$ edge or $K$ pre-edge    
are not directly affected by disorder~\cite{Frenkel2006}.  
   
Our finding clearly shows the limits of the common assumption that 
temperature/doping- and pressure-driven MIT in V$_2$O$_3$ can be 
equivalently described within the same phase 
diagram~\cite{McWhan1970},\cite{footnote2}.  Indeed, the two different 
PM electronic structures that we observed reflect different mechanisms 
driving the MIT along different pathways. In the 
doping-driven MIT, the metallic phase is characterized by an increased 
occupation of the $a_{1g}$ electrons indicating a reduced ``effective 
crystal-field-splitting'' as the main driving mechanism towards 
metallicity~\cite{Keller2004, Poteryaev2007}, related to  
the jump of the lattice parameter $c/a$ 
(1.4\%) at the MIT~\cite{McWhan1970}. In contrast, when pressure is 
applied the $a_{1g}$ occupation remains basically unchanged, so that 
this metallic phase seems to originate rather from an increased 
bandwidth, without any relevant changes of the orbital splitting. The 
smaller $c/a$ jump observed under pressure (0.7 \%) corroborates our 
analysis.   
 
In conclusion, doping, temperature and pressure are shown to act
differently on the interplay between electron correlations and
crystal field, so that states previously considered to be equivalent
metals are actually different. We believe this
finding may apply to many other strongly correlated systems presenting
metal-insulator transitions, with pressure opening inequivalent
pathways through their phase diagrams.   
 
%\begin{acknowledgements} 
We thank E. Tosatti for stimulating discussions and acknowledge a BQR 
of the Universit\'{e} Paris-Sud, the ``Triangle de la Physique'', 
the Austrian Science Fund (FWF, science college W004), and the EU
research network RP7 MONAMI for financial support.  
%\end{acknowledgements} 
 
%\bibliography{e:/TeX/current_jab,extra}   

\begin{thebibliography}{28}
\expandafter\ifx\csname natexlab\endcsname\relax\def\natexlab#1{#1}\fi
\expandafter\ifx\csname bibnamefont\endcsname\relax
  \def\bibnamefont#1{#1}\fi
\expandafter\ifx\csname bibfnamefont\endcsname\relax
  \def\bibfnamefont#1{#1}\fi
\expandafter\ifx\csname citenamefont\endcsname\relax
  \def\citenamefont#1{#1}\fi
\expandafter\ifx\csname url\endcsname\relax
  \def\url#1{\texttt{#1}}\fi
\expandafter\ifx\csname urlprefix\endcsname\relax\def\urlprefix{URL }\fi
\providecommand{\bibinfo}[2]{#2}
\providecommand{\eprint}[2][]{\url{#2}}

\bibitem[{\citenamefont{Imada et~al.}(1998)\citenamefont{Imada, Fujimori, and
  Tokura}}]{Imada1998}
\bibinfo{author}{\bibfnamefont{M.}~\bibnamefont{Imada}},
  \bibinfo{author}{\bibfnamefont{A.}~\bibnamefont{Fujimori}}, \bibnamefont{and}
  \bibinfo{author}{\bibfnamefont{Y.}~\bibnamefont{Tokura}},
  \bibinfo{journal}{Rev. Mod. Phys.} \textbf{\bibinfo{volume}{70}},
  \bibinfo{pages}{1039} (\bibinfo{year}{1998}).

\bibitem[{\citenamefont{Dagotto}(2005)}]{Dagotto2005}
\bibinfo{author}{\bibfnamefont{E.}~\bibnamefont{Dagotto}},
  \bibinfo{journal}{Science} \textbf{\bibinfo{volume}{309}},
  \bibinfo{pages}{257} (\bibinfo{year}{2005}).

\bibitem[{\citenamefont{Georges et~al.}(1996)\citenamefont{Georges, Kotliar,
  Krauth, and Rozenberg}}]{Georges1996}
\bibinfo{author}{\bibfnamefont{A.}~\bibnamefont{Georges}},
  \bibinfo{author}{\bibfnamefont{G.}~\bibnamefont{Kotliar}},
  \bibinfo{author}{\bibfnamefont{W.}~\bibnamefont{Krauth}}, \bibnamefont{and}
  \bibinfo{author}{\bibfnamefont{M.~J.} \bibnamefont{Rozenberg}},
  \bibinfo{journal}{Rev. Mod. Phys.} \textbf{\bibinfo{volume}{68}},
  \bibinfo{pages}{13} (\bibinfo{year}{1996}).

\bibitem[{\citenamefont{McWhan et~al.}(1973)\citenamefont{McWhan, Menth,
  Remeika, Brinkman, and Rice}}]{McWhan1973}
\bibinfo{author}{\bibfnamefont{D.~B.} \bibnamefont{McWhan}},
  \bibinfo{author}{\bibfnamefont{A.}~\bibnamefont{Menth}},
  \bibinfo{author}{\bibfnamefont{J.~P.} \bibnamefont{Remeika}},
  \bibinfo{author}{\bibfnamefont{W.~F.} \bibnamefont{Brinkman}},
  \bibnamefont{and} \bibinfo{author}{\bibfnamefont{T.~M.} \bibnamefont{Rice}},
  \bibinfo{journal}{Phys. Rev. B} \textbf{\bibinfo{volume}{7}},
  \bibinfo{pages}{1920} (\bibinfo{year}{1973}).

\bibitem[{\citenamefont{Castellani et~al.}(1978)\citenamefont{Castellani,
  Natoli, and Ranninger}}]{Castellani1978}
\bibinfo{author}{\bibfnamefont{C.}~\bibnamefont{Castellani}},
  \bibinfo{author}{\bibfnamefont{C.~R.} \bibnamefont{Natoli}},
  \bibnamefont{and}
  \bibinfo{author}{\bibfnamefont{J.}~\bibnamefont{Ranninger}},
  \bibinfo{journal}{Phys. Rev. B} \textbf{\bibinfo{volume}{18}},
  \bibinfo{pages}{4945} (\bibinfo{year}{1978}).

\bibitem[{\citenamefont{Held et~al.}(2001)\citenamefont{Held, Nekrasov,
  Bl{\"u}mer, Anisimov, and Vollhardt}}]{Held2001}
\bibinfo{author}{\bibfnamefont{K.}~\bibnamefont{Held}},
  \bibinfo{author}{\bibfnamefont{I.}~\bibnamefont{Nekrasov}},
  \bibinfo{author}{\bibfnamefont{N.}~\bibnamefont{Bl{\"u}mer}},
  \bibinfo{author}{\bibfnamefont{V.}~\bibnamefont{Anisimov}}, \bibnamefont{and}
  \bibinfo{author}{\bibfnamefont{D.}~\bibnamefont{Vollhardt}},
  \bibinfo{journal}{Int. J. Mod. Phys. B} \textbf{\bibinfo{volume}{15}},
  \bibinfo{pages}{2611} (\bibinfo{year}{2001}).

\bibitem[{\citenamefont{Elfimov et~al.}(2002)\citenamefont{Elfimov, Skorikov,
  Anisimov, and Sawatzky}}]{Elfimov2002a}
\bibinfo{author}{\bibfnamefont{I.~S.} \bibnamefont{Elfimov}},
  \bibinfo{author}{\bibfnamefont{N.~A.} \bibnamefont{Skorikov}},
  \bibinfo{author}{\bibfnamefont{V.~I.} \bibnamefont{Anisimov}},
  \bibnamefont{and} \bibinfo{author}{\bibfnamefont{G.~A.}
  \bibnamefont{Sawatzky}}, \bibinfo{journal}{Phys. Rev. Lett.}
  \textbf{\bibinfo{volume}{88}}, \bibinfo{pages}{015504}
  (\bibinfo{year}{2002}).

\bibitem[{\citenamefont{Keller et~al.}(2004)\citenamefont{Keller, Held, Eyert,
  Vollhardt, and Anisimov}}]{Keller2004}
\bibinfo{author}{\bibfnamefont{G.}~\bibnamefont{Keller}},
  \bibinfo{author}{\bibfnamefont{K.}~\bibnamefont{Held}},
  \bibinfo{author}{\bibfnamefont{V.}~\bibnamefont{Eyert}},
  \bibinfo{author}{\bibfnamefont{D.}~\bibnamefont{Vollhardt}},
  \bibnamefont{and} \bibinfo{author}{\bibfnamefont{V.~I.}
  \bibnamefont{Anisimov}}, \bibinfo{journal}{Phys. Rev. B}
  \textbf{\bibinfo{volume}{70}}, \bibinfo{pages}{205116}
  (\bibinfo{year}{2004}).

\bibitem[{\citenamefont{Poteryaev et~al.}(2007)\citenamefont{Poteryaev,
  Tomczak, Biermann, Georges, Lichtenstein, Rubtsov, Saha-Dasgupta, and
  Andersen}}]{Poteryaev2007}
\bibinfo{author}{\bibfnamefont{A.~I.} \bibnamefont{Poteryaev}},
  \bibinfo{author}{\bibfnamefont{J.~M.} \bibnamefont{Tomczak}},
  \bibinfo{author}{\bibfnamefont{S.}~\bibnamefont{Biermann}},
  \bibinfo{author}{\bibfnamefont{A.}~\bibnamefont{Georges}},
  \bibinfo{author}{\bibfnamefont{A.~I.} \bibnamefont{Lichtenstein}},
  \bibinfo{author}{\bibfnamefont{A.~N.} \bibnamefont{Rubtsov}},
  \bibinfo{author}{\bibfnamefont{T.}~\bibnamefont{Saha-Dasgupta}},
  \bibnamefont{and} \bibinfo{author}{\bibfnamefont{O.~K.}
  \bibnamefont{Andersen}}, \bibinfo{journal}{Phys. Rev. B}
  \textbf{\bibinfo{volume}{76}}, \bibinfo{pages}{085127}
  (\bibinfo{year}{2007}).

\bibitem[{\citenamefont{Laad et~al.}(2006)\citenamefont{Laad, Craco, and
  Muller-Hartmann}}]{Laad2006}
\bibinfo{author}{\bibfnamefont{M.~S.} \bibnamefont{Laad}},
  \bibinfo{author}{\bibfnamefont{L.}~\bibnamefont{Craco}}, \bibnamefont{and}
  \bibinfo{author}{\bibfnamefont{E.}~\bibnamefont{Muller-Hartmann}},
  \bibinfo{journal}{Phys. Rev. B} \textbf{\bibinfo{volume}{73}},
  \bibinfo{eid}{045109} (pages~\bibinfo{numpages}{15}) (\bibinfo{year}{2006}).

\bibitem[{\citenamefont{Tanaka}(2002)}]{Tanaka2002}
\bibinfo{author}{\bibfnamefont{A.}~\bibnamefont{Tanaka}}, \bibinfo{journal}{J.
  Phys. Soc. Jpn.} \textbf{\bibinfo{volume}{71}}, \bibinfo{pages}{1091}
  (\bibinfo{year}{2002}).

\bibitem[{\citenamefont{Limelette et~al.}(2003)\citenamefont{Limelette,
  Georges, Jerome, Wzietek, Metcalf, and Honig}}]{Limelette2003}
\bibinfo{author}{\bibfnamefont{P.}~\bibnamefont{Limelette}},
  \bibinfo{author}{\bibfnamefont{A.}~\bibnamefont{Georges}},
  \bibinfo{author}{\bibfnamefont{D.}~\bibnamefont{Jerome}},
  \bibinfo{author}{\bibfnamefont{P.}~\bibnamefont{Wzietek}},
  \bibinfo{author}{\bibfnamefont{P.}~\bibnamefont{Metcalf}}, \bibnamefont{and}
  \bibinfo{author}{\bibfnamefont{J.~M.} \bibnamefont{Honig}},
  \bibinfo{journal}{Science} \textbf{\bibinfo{volume}{302}},
  \bibinfo{pages}{89} (\bibinfo{year}{2003}).

\bibitem[{\citenamefont{Mo et~al.}(2003)\citenamefont{Mo, Denlinger, Kim, Park,
  Allen, Sekiyama, Yamasaki, Kadono, Suga, Saitoh et~al.}}]{Mo2003}
\bibinfo{author}{\bibfnamefont{S.-K.} \bibnamefont{Mo}},
  \bibinfo{author}{\bibfnamefont{J.~D.} \bibnamefont{Denlinger}},
  \bibinfo{author}{\bibfnamefont{H.-D.} \bibnamefont{Kim}},
  \bibinfo{author}{\bibfnamefont{J.-H.} \bibnamefont{Park}},
  \bibinfo{author}{\bibfnamefont{J.~W.} \bibnamefont{Allen}},
  \bibinfo{author}{\bibfnamefont{A.}~\bibnamefont{Sekiyama}},
  \bibinfo{author}{\bibfnamefont{A.}~\bibnamefont{Yamasaki}},
  \bibinfo{author}{\bibfnamefont{K.}~\bibnamefont{Kadono}},
  \bibinfo{author}{\bibfnamefont{S.}~\bibnamefont{Suga}},
  \bibinfo{author}{\bibfnamefont{Y.}~\bibnamefont{Saitoh}},
  \bibnamefont{et~al.}, \bibinfo{journal}{Phys. Rev. Lett.}
  \textbf{\bibinfo{volume}{90}}, \bibinfo{pages}{186403}
  (\bibinfo{year}{2003}).

\bibitem[{\citenamefont{Mo et~al.}(2006)\citenamefont{Mo, Kim, Denlinger,
  Allen, Park, Sekiyama, Yamasaki, Suga, Saitoh, Muro et~al.}}]{Mo2006}
\bibinfo{author}{\bibfnamefont{S.-K.} \bibnamefont{Mo}},
  \bibinfo{author}{\bibfnamefont{H.-D.} \bibnamefont{Kim}},
  \bibinfo{author}{\bibfnamefont{J.~D.} \bibnamefont{Denlinger}},
  \bibinfo{author}{\bibfnamefont{J.~W.} \bibnamefont{Allen}},
  \bibinfo{author}{\bibfnamefont{J.-H.} \bibnamefont{Park}},
  \bibinfo{author}{\bibfnamefont{A.}~\bibnamefont{Sekiyama}},
  \bibinfo{author}{\bibfnamefont{A.}~\bibnamefont{Yamasaki}},
  \bibinfo{author}{\bibfnamefont{S.}~\bibnamefont{Suga}},
  \bibinfo{author}{\bibfnamefont{Y.}~\bibnamefont{Saitoh}},
  \bibinfo{author}{\bibfnamefont{T.}~\bibnamefont{Muro}}, \bibnamefont{et~al.},
  \bibinfo{journal}{Phys. Rev. B} \textbf{\bibinfo{volume}{74}},
  \bibinfo{pages}{165101} (\bibinfo{year}{2006}).

\bibitem[{\citenamefont{Rodolakis et~al.}(2009)\citenamefont{Rodolakis,
  Mansart, Papalazarou, Gorovikov, Vilmercati, Petaccia, Goldoni, Rueff, Lupi,
  Metcalf and Marsi}}]{Rodolakis2009}
\bibinfo{author}{\bibfnamefont{F.}~\bibnamefont{Rodolakis}},
  \bibinfo{author}{\bibfnamefont{B.}~\bibnamefont{Mansart}},
  \bibinfo{author}{\bibfnamefont{E.}~\bibnamefont{Papalazarou}},
  \bibinfo{author}{\bibfnamefont{S.}~\bibnamefont{Gorovikov}},
  \bibinfo{author}{\bibfnamefont{P.}~\bibnamefont{Vilmercati}},
  \bibinfo{author}{\bibfnamefont{L.}~\bibnamefont{Petaccia}},
  \bibinfo{author}{\bibfnamefont{A.}~\bibnamefont{Goldoni}},
  \bibinfo{author}{\bibfnamefont{J.-P.} \bibnamefont{Rueff}},
  \bibinfo{author}{\bibfnamefont{S.}~\bibnamefont{Lupi}},
  \bibinfo{author}{\bibfnamefont{P.}~\bibnamefont{Metcalf}},
  \bibnamefont{and} \bibinfo{author}{\bibfnamefont{M.}~\bibnamefont{Marsi}}, 
  \bibinfo{journal}{Phys. Rev. Lett.}
  \textbf{\bibinfo{volume}{102}}, \bibinfo{pages}{066805}
  (\bibinfo{year}{2009}).

\bibitem[{\citenamefont{Park et~al.}(2000)\citenamefont{Park, Tjeng, Tanaka,
  Allen, Chen, Metcalf, Honig, de~Groot, and Sawatzky}}]{Park2000}
\bibinfo{author}{\bibfnamefont{J.-H.} \bibnamefont{Park}},
  \bibinfo{author}{\bibfnamefont{L.~H.} \bibnamefont{Tjeng}},
  \bibinfo{author}{\bibfnamefont{A.}~\bibnamefont{Tanaka}},
  \bibinfo{author}{\bibfnamefont{J.~W.} \bibnamefont{Allen}},
  \bibinfo{author}{\bibfnamefont{C.~T.} \bibnamefont{Chen}},
  \bibinfo{author}{\bibfnamefont{P.}~\bibnamefont{Metcalf}},
  \bibinfo{author}{\bibfnamefont{J.~M.} \bibnamefont{Honig}},
  \bibinfo{author}{\bibfnamefont{F.~M.~F.} \bibnamefont{de~Groot}},
  \bibnamefont{and} \bibinfo{author}{\bibfnamefont{G.~A.}
  \bibnamefont{Sawatzky}}, \bibinfo{journal}{Phys. Rev. B}
  \textbf{\bibinfo{volume}{61}}, \bibinfo{pages}{11506} (\bibinfo{year}{2000}).

\bibitem[{\citenamefont{Held}(2007)}]{HeldAdvPhys}
\bibinfo{author}{\bibfnamefont{K.}~\bibnamefont{Held}}, \bibinfo{journal}{Adv.
  in Phys.} \textbf{\bibinfo{volume}{56}}, \bibinfo{pages}{829}
  (\bibinfo{year}{2007}).

\bibitem[{\citenamefont{de~Groot}(2001)}]{Groot2001}
\bibinfo{author}{\bibfnamefont{F.~M.~F.} \bibnamefont{de~Groot}},
  \bibinfo{journal}{Chem. Rev.} \textbf{\bibinfo{volume}{101}},
  \bibinfo{pages}{1779} (\bibinfo{year}{2001}).

\bibitem[{\citenamefont{Bianconi and Natoli}(1978)}]{Bianconi1978}
\bibinfo{author}{\bibfnamefont{A.}~\bibnamefont{Bianconi}} \bibnamefont{and}
  \bibinfo{author}{\bibfnamefont{C.~R.} \bibnamefont{Natoli}},
  \bibinfo{journal}{Solid State Commun.} \textbf{\bibinfo{volume}{27}},
  \bibinfo{pages}{1177} (\bibinfo{year}{1978}).

\bibitem[{\citenamefont{Goulon et~al.}(2000)\citenamefont{Goulon, Rogalev,
  Goulon-Ginet, Benayoun, Paolasini, Brouder, Malgrange, and
  Metcalf}}]{Goulon2000}
\bibinfo{author}{\bibfnamefont{J.}~\bibnamefont{Goulon}},
  \bibinfo{author}{\bibfnamefont{A.}~\bibnamefont{Rogalev}},
  \bibinfo{author}{\bibfnamefont{C.}~\bibnamefont{Goulon-Ginet}},
  \bibinfo{author}{\bibfnamefont{G.}~\bibnamefont{Benayoun}},
  \bibinfo{author}{\bibfnamefont{L.}~\bibnamefont{Paolasini}},
  \bibinfo{author}{\bibfnamefont{C.}~\bibnamefont{Brouder}},
  \bibinfo{author}{\bibfnamefont{C.}~\bibnamefont{Malgrange}},
  \bibnamefont{and} \bibinfo{author}{\bibfnamefont{P.~A.}
  \bibnamefont{Metcalf}}, \bibinfo{journal}{Phys. Rev. Lett.}
  \textbf{\bibinfo{volume}{85}}, \bibinfo{pages}{4385} (\bibinfo{year}{2000}).


\bibitem[{\citenamefont{T. Saha-Dasgupta et~al.}(2007)\citenamefont{Saha-Dasgupta,Andersen,Nuss, Poteryaev,Georges,Lichtenstein}}]{paperI}
\bibinfo{author}{\bibfnamefont{T.}~\bibnamefont{Saha-Dasgupta}},
\bibinfo{author}{\bibfnamefont{O.~K.}~\bibnamefont{Andersen}},
\bibinfo{author}{\bibfnamefont{J.~}~\bibnamefont{Nuss}},
\bibinfo{author}{\bibfnamefont{A.~I.} \bibnamefont{Poteryaev}},
\bibinfo{author}{\bibfnamefont{A.}~\bibnamefont{Georges}}, \bibnamefont{and} 
\bibinfo{author}{\bibfnamefont{A.~I.} \bibnamefont{Lichtenstein}},\bibinfo{journal}{ arXiv:0907.2841 (preprint 2009)}.



\bibitem[{\citenamefont{Gougoussis et~al.}(2009)\citenamefont{Gougoussis,
  Calandra, Seitsonen, Brouder, Shukla, and Mauri}}]{Gougoussis2009}
\bibinfo{author}{\bibfnamefont{C.}~\bibnamefont{Gougoussis}},
  \bibinfo{author}{\bibfnamefont{M.}~\bibnamefont{Calandra}},
  \bibinfo{author}{\bibfnamefont{A.}~\bibnamefont{Seitsonen}},
  \bibinfo{author}{\bibfnamefont{C.}~\bibnamefont{Brouder}},
  \bibinfo{author}{\bibfnamefont{A.}~\bibnamefont{Shukla}}, \bibnamefont{and}
  \bibinfo{author}{\bibfnamefont{F.}~\bibnamefont{Mauri}},
  \bibinfo{journal}{Phys. Rev. B} \textbf{\bibinfo{volume}{79}},
  \bibinfo{pages}{045118} (\bibinfo{year}{2009}).

\bibitem{footnote}{Constrained LDA/LDA+U estimates of $J$ (0.93eV in
  \cite{Solovyev}) need to be reduced to $J \sim 0.7$eV to reproduce the
  constrained LDA energy splitting with the DMFT Hamiltonian
  \cite{HeldAdvPhys}. Moreover, the value of U adopted for LDA+U ($2.8$eV) has
  to be assumed smaller than that for LDA+DMFT to compensate deficiencies of
  LDA+U \cite{Sangiovanni}}.

\bibitem[{\citenamefont{Filipponi}(2000)}]{Filipponi2000}
\bibinfo{author}{\bibfnamefont{A.}~\bibnamefont{Filipponi}},
  \bibinfo{journal}{J. Phys. B: At. Mol. Opt. Phys.}
  \textbf{\bibinfo{volume}{33}}, \bibinfo{pages}{2835} (\bibinfo{year}{2000}).

\bibitem[{\citenamefont{McWhan and Remeika}(1970)}]{McWhan1970}
\bibinfo{author}{\bibfnamefont{D.~B.} \bibnamefont{McWhan}} \bibnamefont{and}
  \bibinfo{author}{\bibfnamefont{J.~P.} \bibnamefont{Remeika}},
  \bibinfo{journal}{Phys. Rev. B} \textbf{\bibinfo{volume}{2}},
  \bibinfo{pages}{3734} (\bibinfo{year}{1970}).

\bibitem[{\citenamefont{Frenkel et~al.}(2006)\citenamefont{Frenkel, Pease,
  Budnick, Metcalf, Stern, Shanthakumar, and Huang}}]{Frenkel2006}
\bibinfo{author}{\bibfnamefont{A.~I.} \bibnamefont{Frenkel}},
  \bibinfo{author}{\bibfnamefont{D.~M.} \bibnamefont{Pease}},
  \bibinfo{author}{\bibfnamefont{J.~I.} \bibnamefont{Budnick}},
  \bibinfo{author}{\bibfnamefont{P.}~\bibnamefont{Metcalf}},
  \bibinfo{author}{\bibfnamefont{E.~A.} \bibnamefont{Stern}},
  \bibinfo{author}{\bibfnamefont{P.}~\bibnamefont{Shanthakumar}},
  \bibnamefont{and} \bibinfo{author}{\bibfnamefont{T.}~\bibnamefont{Huang}},
  \bibinfo{journal}{Phys. Rev. Lett.} \textbf{\bibinfo{volume}{97}},
  \bibinfo{pages}{195502} (\bibinfo{year}{2006}).

\bibitem{footnote2}{An early version of the phase diagram (Fig.~15 in
  Ref.~\onlinecite{McWhan1970}) was drawn with a third pressure axis: due to
  the idea of $P$-$x$ equivalence, this was later abandoned.}

\bibitem[{\citenamefont{Solovyev et~al.}(1996)\citenamefont{Solovyev, Hamada,
  and Terakura}}]{Solovyev}
\bibinfo{author}{\bibfnamefont{I.}~\bibnamefont{Solovyev}},
  \bibinfo{author}{\bibfnamefont{N.}~\bibnamefont{Hamada}}, \bibnamefont{and}
  \bibinfo{author}{\bibfnamefont{K.}~\bibnamefont{Terakura}},
  \bibinfo{journal}{Phys. Rev. B} \textbf{\bibinfo{volume}{53}},
  \bibinfo{pages}{7158} (\bibinfo{year}{1996}).

\bibitem[{\citenamefont{Sangiovanni et~al.}(2006)\citenamefont{Sangiovanni,
  Toschi, Koch, Held, Capone, Castellani, Gunnarsson, Mo, Allen, Kim
  et~al.}}]{Sangiovanni}
\bibinfo{author}{\bibfnamefont{G.}~\bibnamefont{Sangiovanni}},
  \bibinfo{author}{\bibfnamefont{A.}~\bibnamefont{Toschi}},
  \bibinfo{author}{\bibfnamefont{E.}~\bibnamefont{Koch}},
  \bibinfo{author}{\bibfnamefont{K.}~\bibnamefont{Held}},
  \bibinfo{author}{\bibfnamefont{M.}~\bibnamefont{Capone}},
  \bibinfo{author}{\bibfnamefont{C.}~\bibnamefont{Castellani}},
  \bibinfo{author}{\bibfnamefont{O.}~\bibnamefont{Gunnarsson}},
  \bibinfo{author}{\bibfnamefont{S.-K.} \bibnamefont{Mo}},
  \bibinfo{author}{\bibfnamefont{J.~W.} \bibnamefont{Allen}},
  \bibinfo{author}{\bibfnamefont{H.-D.} \bibnamefont{Kim}},
  \bibnamefont{et~al.}, \bibinfo{journal}{Phys. Rev. B}
  \textbf{\bibinfo{volume}{73}}, \bibinfo{pages}{205121}
  (\bibinfo{year}{2006}).

\end{thebibliography}
%\bibliographystyle{apsrev}  

\end{document}